# Study of Aqueous Dispersions of Magnetic Nanoparticles by Magnetic and Rheological Measurements

Shalini .M
*Department of Physics, University of Mumbai and UM-DAE Centre for Excellence in Basic Sciences, Mumbai-400098, India*
S. Radha, D. C. Kothari[#]
*Department of Physics, University of Mumbai, Mumbai – 400098, India*

[#]*National Centre for Nanosciences and Nanotechnology, University of Mumbai, Mumbai-400098, India*

**Abstract**

The observed magnetic tunability of light transmission through a ferrofluid can be effectively understood in terms of the inter-particle interaction that can be estimated from the magnetic and rheological properties of these fluids. The present study reports complementary magnetic and rheological measurements of aqueous dispersions of ferrite nanoparticles and a commercial ferrofluid. The room temperature magnetization measured in a SQUID magnetometer up to fields of 1 to 2 Tesla showed superparamagnetic behaviour of the particles and the dispersion with the background signal of the liquid showing a diamagnetic behaviour. The room temperature rheological behaviour in zero magnetic field of the fluids was investigated by measuring the viscosity as a function of shear rate from 1-100 $s^{-1}$. The particle size and the nature of the carrier liquid determine the viscosity and is expected to have an effect on the inter-particle interaction.

Email: radhas@physics.mu.ac.in

Fax : +91-22-26529780

# I. INTRODUCTION

Magnetic nanocolloids have many interesting properties which find applications in sealing, printing targeted drug delivery etc. Aqueous dispersions of SPIONs (SuperParamagnetic Iron Oxide Nanoparticles), specially have been the subject of intense research due to their inherent biocompatibility [1-3]. Generally, these SPIONs are coated with a surfactant to prevent agglomeration that occurs due to presence of competing Van der Waals forces and the magnetic dipole interaction between the particles. Among various nano-materials, magnetic nanocrystals have been widely used in many applications such as in ferrofluids, recording media, electronic, optical and biomedical disciplines. In addition, magnetic colloids have been found to be an ideal platform to study fundamental aspects of magnetism such as magnetic domain formation, models for magnetic structure, superparamagnetic behavior etc.

Magneto-optical effects in ferrofluids have generated a lot of interest due to the tunability of the medium by the magnetic field. The studies on variation of light transmission through a ferrolfluid with different geometries of magnetic field and incident light have been widely reported [4-7]. These effects have been attributed to structures formed inside the medium due to the interaction of the field with the particles. Recently Shalini et al [8] have reported the observation of simultaneous decrease of forward transmission and back reflection from a ferrofluid comprising of magnetite nanoparticles suspended in kerosene, in a transverse magnetic field. As the magnetic field is slowly increased, the fluid goes from disordered to a partially ordered state that is modeled to have chain-like structures with superimposed randomness. The interplay of partial order, disorder and dissipation affects the scattering that explains the observed optical behavior. There is an interest in further exploring this system by complementary studies of physical properties of the carrier liquid and the collective magnetic response. The present work reports the complementary rheological and magnetic property measurements of some ferrofluids.

## II. EXPERIMENTS AND DISCUSSION

The ferrite nanoparticles were prepared by standard wet chemical routes of co-precipitation of $Fe^{2+}$ and $Fe^{3+}$ ions by $NH_4OH$ at temperatures above 60 °C and were later dispersed in double distilled water to form reasonably stable dispersions. The size of the nanoparticles as estimated by Scherrer formula from XRD measurements was in the range 10-28 nm. A PC based pulsed field hysteresis loop tracer was used for observing the hysteresis loops of the powder samples at room temperature [9]. The standard sample was a commercial ferrofluid from Ferrolabs Inc. USA (FLS 300.020). The magnetization measurements at room temperature for the fluid samples were carried out on a SQUID magnetometer (Quantum Design MPMS X7) between fields of 1 to 2 Tesla. The liquid samples were sealed in a micropipette tip, closed on both the sides with Araldite, since these tips were found suitable for sensitive magnetometers due to their very low background signal. Viscosity measurements were carried out using a Rotational Rheometer (Haake Mars Rheometer, Thermo Scientific, Germany). The measurement was done using a Cone/Plate geometry using Cone/35mm/1° at room temperature 25 °C. The samples were tested for a time interval of 600s at a shear rate range of 1-100 $s^{-1}$. The flow curve was obtained by plotting shear stress vs shear rate. The data was further fitted with suitable models based on the flow curve using the Haake rheowin software. In this case, the Oswald De Waale Power law model was adopted. This yielded the viscosity curve that shows the variation of viscosity with respect to shear rate.

Figure 1(a) shows the variation of magnetization, M vs magnetic field, H for magnetite nanoparticles at room temperature, using a loop tracer up to fields of 6kOe. It showed a saturation magnetization of 54.8 emu/g and coercivity of 110 Oe. The standard sample was diluted to 5% concentration by volume using kerosene. Figure 1(b) shows a representative plot of M-H loop of the standard liquid samples from SQUID magnetometer. The background signal from the blank liquid (red curve) shows a diamagnetic behaviour of magnitude much smaller than that of the magnetic fluid (blue curve) (see inset). The magnetic fluid samples showed a saturation magnetization of 20-22 emu/g and $H_c$=12 Oe. However, according to the specifications given by the manufacturers, the saturation magnetization of the stock is about 3-6 emu/g. This was observed in all the standard samples and indicates an increase in saturation magnetization of the ferrofluid dispersions. A similar effect of increased saturation magnetization of dispersed magnetic nanoparticles has been earlier observed [10,11]. The observed M-H curve with large saturation magnetization and low coercivity is a characteristic of small domain-particles that in the superparamagnetic limit exhibit a non-saturating magnetization and an anhysteretic (zero coercivity) behavior. It is evident that the qualitative nature of the hysteresis curve of the standard sample remains unaltered when dispersed in fluid except for the increased saturation magnetization. The magnetic measurement on the present batch of synthesized nanoparticles in dispersion has not been possible as of date, though earlier measurements on similar systems have indicated an increase in magnetization [11]. The higher coercivity of the synthesized nanoparticles compared to the standard one indicates a larger average grain size arising due to the polydispersity.

The viscosity of liquid samples is estimated as the ratio of the shear stress to shear rate using the relation $\tau = \eta \gamma'$, where $\gamma'$ is the shear rate and $\eta$ is the viscosity. A linear relationship of the shear stress with shear rate is ascribed as Newtonian behavior. A non-linear flow curve indicates a non-Newtonian fluid where the viscosity changes with shear rate as against the Newtonian fluid where the viscosity is constant. If the viscosity decreases with increasing shear rate, the sample is said to be showing shear thinning behavior whereas if it increases with increases with increasing shear rate, it is said to show shear thickening behavior. Mathematically, this relationship between shear rate and shear stress is given as $\tau = K \gamma'^n$ using the Ostwald-de Waele model. Here n is the flow index. If $n > 1$, the fluid is said to show dilatants or shear thickening behavior while if n<1, it is said to show pseudo-plastic or shear-thinning behavior. K is called the consistency factor that is a measure of the apparent viscosity.

The values of viscosity and the flow index arrived at fitting the data of the samples to these models are tabulated in Table1.

The results of the viscosity study of ferrofluids comprising of magnetite nanoparticles dispersed in water (FF2) and kerosene (standard; FF1) are shown in Figs. 2 to 4. Fig 2(b) shows the viscosity curve of blank liquid - kerosene. It can be seen that the base fluid shows Newtonian behaviour at room temperature. The viscosity measured at room temperature of water is 1.0 cP (mPa.s) and for Kerosene is 1.6 cP (mPa.s). In figure 3(b), the viscosity of the standard sample FF1 (MNPs dispersed in kerosene) is constant over shear rate throughout, indicating a Newtonian flow behavior. Fig. 4(a) shows the flow curve for the synthesized sample. At low shear rates, the flow curve for the aqueous dispersion (MNPs dispersed in water) shows a deviation from linearity. This is reflected in the viscosity curve, where it shows a sudden increase at lower shear rates and tends to saturation at higher shear rates. This effect is attributed to a non-Newtonian behavior.

There have been several studies on ferrofluids to understand the magnetic field dependent changes of viscosity by magnetorheological methods [12,13]. These studies correlate the observed magneto-viscous effects [14] with the formation of chain-like clusters of magnetic nanoparticles under certain external conditions, like the applied field, the size and shape of the particles, the solid content of the particles in the fluid and the nature of the fluid. Zubarev et al [15] have pointed out that the polydispersity of the particles also play an important role in the rheological properties of the ferrofluid. The observed difference in viscosity behavior in our samples presumably arises due to the effect of the varying particle sizes in the synthesized ferrofluids as compared to the commercial one, as is also evident from the higher coercivity in the magnetization measurements. The effect of the carrier fluid is not expected to be as significant since both water and kerosene, showed comparable viscosity values and a Newtonian fluid behavior at normal shear rates. The theoretical models for interaction of the magnetic particles in fluids provide an estimate for the interaction parameter that depends directly on the value of the saturation magnetization [15]. Thus it is important to have quantitative magnetization measurements of the magnetic fluids to understand the magneto-rheological behavior.

## II. CONCLUSION

The present study of magnetization and rheology on a set of ferrofluid samples show interesting effects. However, it is necessary to carry out further systematic quantitative measurements on these fluids to understand the observed enhancement in saturation magnetization as well as the variation in viscosity. The nature of the carrier fluid is to be further investigated by studying the field dependent rheological properties. This combined study is expected to provide estimates for the interparticle interaction that leads to the formation of internal structures in the ferrofluids under applied fields. These in turn will enable designing magnetic fluids for several desirable applications.

## ACKNOWLEDGEMENT

The authors would like to thank Prof. A. K. Nigam (TIFR) and his lab staff for XRD and SQUID measurements and Mr Mahesh Samant (ex-CNNUM) for rheology measurements. We also extend our thanks to Prof. Hema Ramachandran (RRI) for providing the commercial ferrofluid samples. Two of the authors (SM and RS) acknowledge Prof. R. Nagarajan (UM-DAE CBS) for valuable discussion and support.

**Table 1 Viscosity and Flow behaviour index (n) for the liquids estimated from various models**

| Sample | Model | η (Pas) | K (consistency factor) | n (flow index) |
|---|---|---|---|---|
| Water | Newtonian | 0.98 | -- | 1 |
| Kerosene | Newtonian | 1.6 | -- | 1 |
| FF 1 (MNP+Kerosene) | Ostwald De waale | 1.5 | -- | 1 |
| FF 2 (MNP-Water) | Ostwald De waale (Non Newtonian) | -- | 0.744 | 1.175 |

**Figure Captions**

Fig. 1(a) M-H loop tracer for magnetite powder from pulsed field loop tracer

Fig. 1(b) Room temperature M-H loop of standard FF (5% volume concentration in kerosene) from SQUID. The red curve shows the background signal of carrier liquid. Inset shows the data at lower fields.

Fig. 2(a)-2(b) Viscosity curve and Flow curve for base fluid Kerosene

Fig. 3(a)-3(b) Viscosity curve and Flow curve of standard sample (MNP in kerosene)

Fig. 4(a) -4(b) Viscosity curve and Flow curve of synthesized sample (MNP in water)

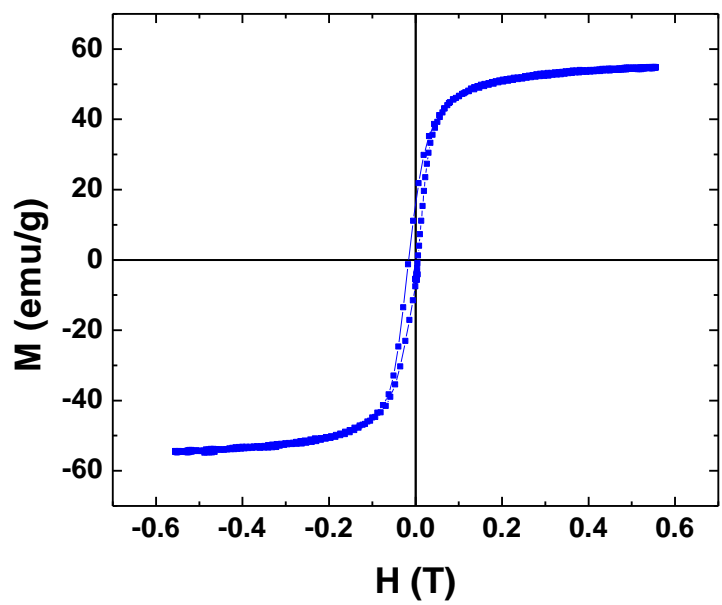

Fig. 1(a)

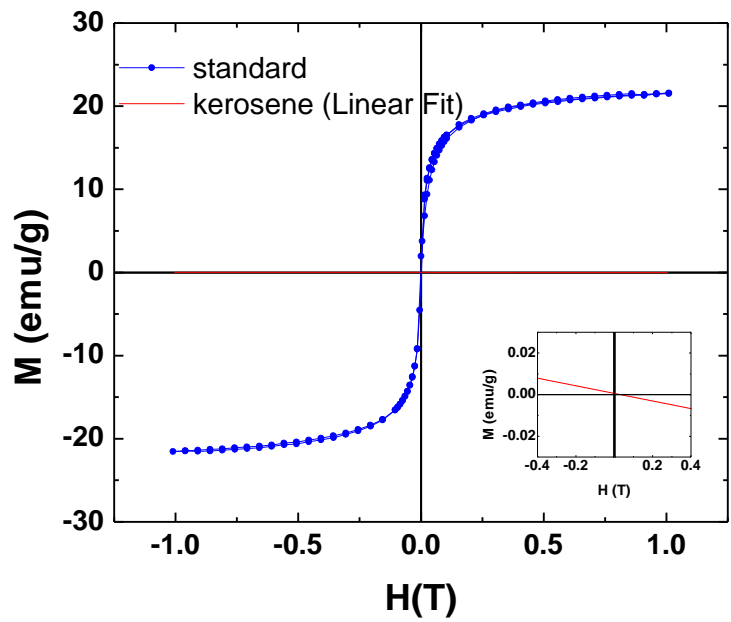

Fig. 1(b)

| Kerosene ||
|---|---|
| 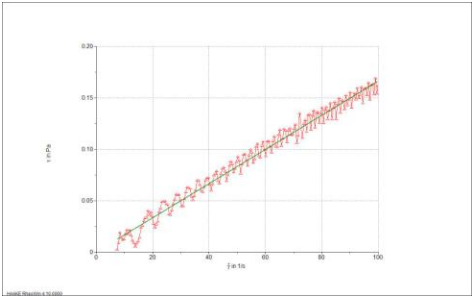 | 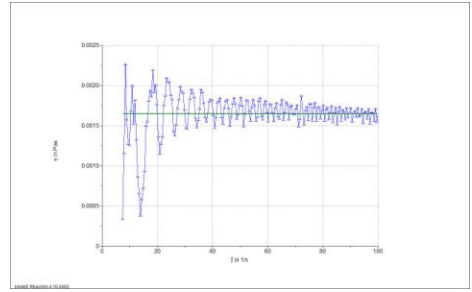 |
| Fig 2(a)- Flow curve | Fig 2(b) Viscosity Curve |
| 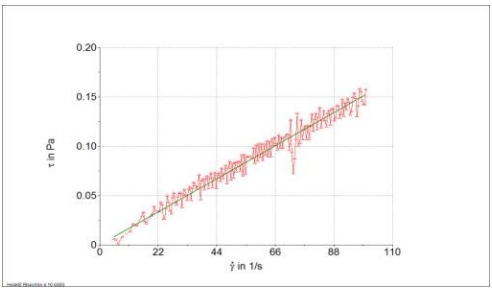 | 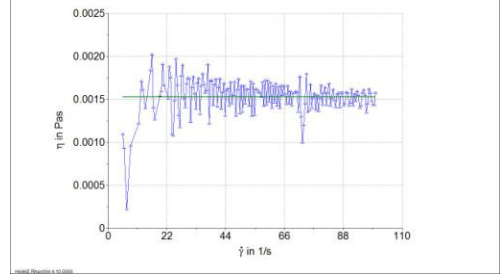 |
| Fig 3(a) Flow curve | Fig 3(b) Viscosity Curve |
| FF 1 (MNP-kerosene) standard sample ||
| 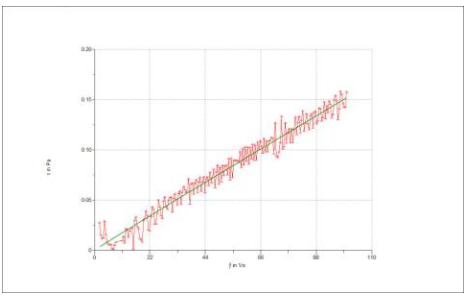 | 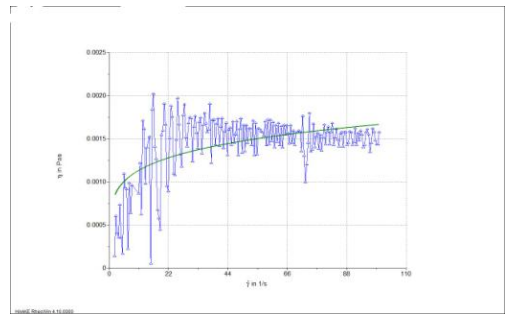 |
| Fig 4(a) Flow curve | Fig 4(b) Viscosity Curve |
| FF 2 (MNP-Water) sample ||